\documentclass[a4paper,11pt]{article}
\pdfoutput=1 

\usepackage{jheppub} 
\usepackage[T1]{fontenc} 
\usepackage{amsmath,amssymb}
\usepackage{dsfont}
\usepackage{upgreek}
\usepackage{slashed}
\usepackage{appendix}
\usepackage{tabu}
\usepackage{physics}
\usepackage{xcolor}

\usepackage{bbm}
\usepackage{mathrsfs}
\usepackage{multirow}
\usepackage{relsize}
\usepackage{caption}
\usepackage{cancel}
\usepackage{graphics,graphicx}
\usepackage{float}
\usepackage{lipsum}
\usepackage{comment}
\usepackage{enumitem}
\numberwithin{equation}{section}
\def\beq{\begin{align}}
\def\eeq{\end{align}}
\newcommand{\bi}{\begin{itemize}}
\newcommand{\ei}{\end{itemize}}
\newcommand{\ben}{\begin{enumerate}}
\newcommand{\een}{\end{enumerate}}
\newcommand{\be}{\begin{equation}}
\newcommand{\ee}{\end{equation}}
\newcommand{\bea}{\begin{eqnarray}}
\newcommand{\eea}{\end{eqnarray}}

\def\del{\partial}
\def\vo{\mathcal{V}}
\usepackage{tikz}
\usepackage{subcaption}

\def\del{\partial}
\def\vo{\mathcal{V}}
\def\pref#1{(\ref{#1})}

\title{Joint Statistics of Cosmological Constant and SUSY Breaking in Flux Vacua with Nilpotent Goldstino}
\author[a,b]{Michele Cicoli,}
\author[a,b]{Matteo Licheri,}
\author[c]{Anshuman Maharana,}
\author[c]{Kajal Singh,}
\author[d]{Kuver Sinha}

\affiliation[a]{\footnotesize Dipartimento di Fisica e Astronomia, Universit\`a di Bologna, via Irnerio 46, 40126 Bologna, Italy}
\affiliation[b]{\footnotesize INFN, Sezione di Bologna, viale Berti Pichat 6/2, 40127 Bologna, Italy}
\affiliation[c]{\footnotesize Harish-Chandra Research Institute, A CI of Homi Bhabha National Institute, Allahabad 211019, India}
\affiliation[d]{\footnotesize Department of Physics and Astronomy, University of Oklahoma, Norman, OK 73019, USA}

\emailAdd{michele.cicoli@unibo.it}
\emailAdd{matteo.licheri@unibo.it}
\emailAdd{anshumanmaharana@hri.res.in}
\emailAdd{kajalsingh@hri.res.in}
\emailAdd{kuversinha@ou.edu}

\abstract{We obtain the joint distribution of the gravitino mass and the cosmological constant
in KKLT and LVS models with anti-D3 brane uplifting described via the nilpotent goldstino formalism. Moduli stabilisation (of both complex structure and
K\"ahler moduli) is incorporated so that we sample only over points corresponding to vacua.
Our key inputs are the distributions of the flux superpotential, the string coupling and the hierarchies of warped throats.
In the limit of zero cosmological constant, we find that both in KKLT and LVS the distributions are tilted favourably towards lower scales of supersymmetry breaking.}

\begin{document} 
\maketitle
\flushbottom

\section{Introduction}

Phenomenologically attractive string compactifications can be obtained by turning on background fluxes \cite{Michelson:1996pn, Dasgupta:1999ss, Gukov:1999ya, Curio:2000sc, Giddings:2001yu}. In the type IIB setting, the 
effect of fluxes is to stabilise the complex structure moduli and axio-dilaton \cite{Giddings:2001yu}. Furthermore, in type IIB there are various scenarios for the stabilisation of the K\"ahler moduli
\cite{Kachru:2003aw, Balasubramanian:2005zx,vonGersdorff:2005bf, Berg:2005yu, Westphal:2006tn, Cicoli:2008va, Cicoli:2012fh, Gallego:2017dvd, Cicoli:2016chb, Antoniadis:2018hqy, AbdusSalam:2020ywo, Cicoli:2015ylx}. This has made
type IIB flux compactifications the setting for various detailed phenomenological explorations.

The introduction of  fluxes also leads to the possibility of a large multitude of solutions. Apart from construction of  detailed models, string phenomenology
involves developing an understanding of the broad properties of vacua. The later has motivated the statistical approach to string phenomenology \cite{Douglas:2003um, Cole:2019enn, Ashok:2003gk, Denef:2004ze, Denef:2004dm, Denef:2004cf} (see also \cite{Marchesano:2004yn, Dienes:2006ut, Gmeiner:2005vz, Douglas:2006xy, Acharya:2005ez, Giryavets:2004zr, Misra:2004ky, Conlon:2004ds, DeWolfe:2005gy, Hebecker:2006bn, Martinez-Pedrera:2012teo, Halverson:2018cio, Betzler:2019kon, Bena:2021wyr, Krippendorf:2021uxu, Susskind:2004uv, Douglas:2004qg, Dine:2004is, Arkani-Hamed:2005zuc, Kallosh:2004yh, Dine:2005yq, Broeckel:2020fdz, Broeckel:2021dpz, Halverson:2019cmy, Mehta:2020kwu, Carifio:2017bov}).  The distribution of the scale of supersymmetry breaking in the space of string vacua is of course of much interest 
(see \cite{Susskind:2004uv, Douglas:2004qg, Dine:2004is, Arkani-Hamed:2005zuc, Dienes:2006ut, Kallosh:2004yh, Dine:2005yq} for early work in this direction).  The goal of this paper is to study the
joint distribution of the gravitino mass (which sets the scale of supersymmetry breaking in the visible sector in these models) and the cosmological constant in  the two most well developed scenarios for K\"ahler moduli stabilisation in type IIB: KKLT \cite{Kachru:2003aw} and LVS \cite{Balasubramanian:2005zx} models. In addition to the scenario for moduli stabilisation, the quantities of interest also
depend on the uplift sector of the models. This paper will focus on models where the uplift sector is an anti-brane at the tip of a warped throat\footnote{There has been much discussion in
the literature on the (meta)stability  of anti-D3 branes in warped throats, see e.g. \cite{DeWolfe:2008zy,McGuirk:2009xx,Bena:2009xk, Bena:2014jaa,Danielsson:2014yga,Michel:2014lva,Hartnett:2015oda,Bena:2015kia,Cohen-Maldonado:2015ssa,Bertolini:2015hua,Polchinski:2015bea,Cohen-Maldonado:2015lyb, Bena:2018fqc, Dudas:2019pls, Gao:2020xqh, Brennan:2017rbf}.} (as in the construction of \cite{Kachru:2003aw}). 

The key ingredients for our analysis will be moduli stabilisation and a systematic incorporation of the effect of the anti-brane via the nilpotent goldstino formalism. Let us describe them in detail.

\begin{itemize}
\item {\bf{Incorporation of moduli stabilisation (both complex structure and K\"ahler):}} The values of the cosmological constant and the gravitino mass (and other observables) in a vacuum are
determined by the expectation value of the moduli fields. Thus, in order to study the distribution of observables, it is important to incorporate moduli stabilisation and compute the distributions
 sampling only over points corresponding to the minima of the moduli potential. Early statistical analysis of observables incorporated the stabilisation of complex structure moduli; the importance
 of K\"ahler moduli stabilisation was instead emphasised recently in  \cite{Broeckel:2020fdz}. Here, the effect of K\"ahler moduli stabilisation on the distribution of the gravitino mass was studied. It was found that
 this has a significant effect on the statistics. Following  \cite{Broeckel:2020fdz} we will sample only over points corresponding to minima of the moduli potential. 
 
\item {\bf{ Incorporation of the uplift sector}:} In both KKLT and LVS, a crucial contribution  to the cosmological constant comes from the so-called uplift sector. Before
the incorporation of this sector into the effective action, the vacua obtained are necessarily AdS. Furthermore,  the cosmological constant and the gravitino mass
are correlated in these vacua. Before considering the uplift sector, KKLT vacua are supersymmetric. Thus (setting $M_P=1$)
\begin{equation}
\label{kkltintro}
  \hat{V}_{\rm KKLT} = - 3 \hat{m}^2_{3/2} \, ,
\end{equation}
where the hat indicates a quantity computed in the effective field theory before adding the uplift sector. Similarly for LVS vacua
(before uplifting)
\begin{equation}
\label{lvsintro}
  \hat{V}_{\rm LVS} \simeq  -\hat{m}^3_{3/2} \, ,
\end{equation}
The relations (\ref{kkltintro}) and (\ref{lvsintro}) are broken solely due to introduction of the uplift sector. Thus it is crucial to incorporate it while computing the
joint distribution of the cosmological constant and the gravitino mass\footnote{Reference \cite{Broeckel:2020fdz} focused on the gravitino mass distribution
before the inclusion of the uplift sector.}. 

Of course, there are various effects that can lead to an uplift. We shall focus on an anti-D3 brane at the bottom of a warped throat. The reasons for this are the following:
    
\begin{itemize}  
\item {\bf{The nilpotent superfield formalism:}} The nilpotent superfield formalism allows for explicit computation of the effect of an anti-D3 brane in a warped throat.
The effects of the anti-brane are captured by the introduction  of a nilpotent chiral superfield $X$ such that $X^2=0$ (see for instance \cite{Rocek:1978nb,Ivanov:1978mx,Lindstrom:1979kq,Casalbuoni:1988xh, Komargodski:2009rz} and references therein). $X$ has a single propagating component, the Volkov-Akulov goldstino \cite{Volkov:1973ix}, and  supersymmetry  is broken by its F-term. The effective  couplings of such a field $X$ were studied in \cite{Farakos:2013ih,Antoniadis:2014oya,Bergshoeff:2015tra, Dudas:2015eha, Antoniadis:2015ala,Hasegawa:2015bza,Kallosh:2015tea,DallAgata:2015pdd,Schillo:2015ssx,Kallosh:2015pho} and the KKLT uplifting term was reproduced within the supergravity framework in \cite{Bergshoeff:2015jxa,Kallosh:2014wsa,Kallosh:2015nia}. Finally, in \cite{Kallosh:2015nia} explicit string constructions were presented in which an anti-D3-brane at the bottom of a warped throat has only the goldstino as its light degree of freedom,  justifying the use of the nilpotent field $X$ to describe the anti-brane. Soft masses in KKLT and LVS were computed using this framework in \cite{Aparicio:2015psl}.

\item {\bf{Distribution of throat hierarchies:}} For uplift  with anti-D3 branes in warped throats, the magnitude of the uplift term  is set by the hierarchy associated with the warped throat.
Thus  an understanding of the distribution of throat hierarchies is needed to understand the distribution of physical observables in this setting. The distribution of throat hierarchies
has been  studied in detail in \cite{Hebecker:2006bn}. We will make heavy use of these results.
\end{itemize}
\end{itemize}

Our main finding is that the distribution of  the gravitino mass at zero cosmological constant is tilted towards lower values. This is not
the same as the result of \cite{Douglas:2004qg} (see also \cite{Susskind:2004uv}) which  carried out generic estimates on supersymmetry breaking F- and D-terms and concluded that
there is a preference for high scale breaking. Note that this result is different also from the one of \cite{Broeckel:2020fdz} which found a logarithmic preference for high scales of supersymmetry breaking after stabilising the K\"ahler moduli but before uplifting. The difference arises due to the presence of the relations (\ref{kkltintro}) and (\ref{lvsintro}) and the
form of the distribution for throat hierarchies. Our results should not be taken as giving the generic picture for string vacua. In fact, in some
class of models such as \cite{Saltman:2004jh} high scale breaking is essentially in-built. At the same time, we find it very interesting that the best understood
models have  distributions favouring lower masses of the gravitino.

This paper is organised as follows. In Sec. \ref{sec:2review} we review some material that will be needed for our analysis. In Sec. \ref{sec:3joint} we compute the joint distribution for the gravitino mass and cosmological constant in LVS and KKLT models with anti-D3 brane uplifting. We discuss our results and conclude in Sec. \ref{sec:4disc}.  We will closely follow the notation and conventions of  \cite{Aparicio:2015psl}.

\section{Review}\label{sec:2review}

 In this section we review material that will be needed for our analysis. We will touch upon 3 topics: ($i$)
 warped throats in type IIB flux compactifications, ($ii$) the statistical distribution of parameters that appear in the
 effective field theory of the K\"ahler moduli, and ($iii$) the nilpotent goldstino formalism.

\subsection{Warped throats in IIB  flux compactifications}
\label{rwt}

In this subsection we review some aspects of warped throats in flux compactifications that will be useful for this paper. Type IIB flux compactifications have 3-form fluxes (NSNS and RR) threading the 3-cycles of an orientifolded Calabi-Yau (CY). The  back-reaction of fluxes has the effect of generating warping (as in \cite{Randall:1999ee}). The 10D metric takes the form \cite{Giddings:2001yu,Dasgupta:1999ss}
\be
\dd s_{10}^2=e^{2A(y)} \eta_{\mu\nu} \dd{x^\mu} \dd{x^\nu}+e^{-2A(y)} g_{mn} \dd{y^m} \dd{y^n},
\label{eq:warpedmetric1} \ 
\ee
where $e^{2A(y)}\equiv h^{-1/2}(y)$ is the warp factor and $g_{mn}$the CY metric. The warp factor satisfies a Poisson-like equation which is sourced by  3-form fluxes and localised objects carrying D3-charge. For non-vanishing fluxes, the warp factor varies over the compact directions.  The warp factor acts like a redshift factor for the objects localised in the compact directions. In regions where $e^{-2A}$ is large, it can be used to generate hierarchies in physical scales. Regions of large warping arise when fluxes thread the 3-cycle associated with a conifold modulus and its dual cycle. The geometry in this region is close to that of the Klebanov-Strassler (KS) throat \cite{Klebanov:2000hb}.  Our primary interest will be in the regime  in which the warp factor is almost constant over the whole compact space, except for a single throat where the geometry is highly warped.

As pointed out in \cite{Giddings:2005ff}, 
a constant shift of $e^{4A}$ maps solutions of the Poisson equation to solutions, and this freedom is to be identified with 
(a power of) the  volume modulus. Furthermore, a pure scaling of the CY metric $\dd s_{CY}^2$ to a unit-volume fiducial metric $\dd s_{CY_0}^2$, given by $\dd s_{CY}^2 \to \lambda \dd s_{CY}^2$, implies a rescaling of the warp factor $e^{2A} \to \lambda e^{2A}$, and hence has no effect on the physical geometry.  Given this, a useful parametrisation of the 10D geometry is
\be
\dd{s^2_{10}}= \vo^{1/3}\left(e^{-4A_0}+\vo^{2/3}\right)^{-1/2} \dd{s^2_4+}\left(e^{-4A_0}+\vo^{2/3}\right) ^{1/2}\dd{s_{CY_0}^2}\label{eq:warpedmetric2}\:,
\ee
which is equivalent to:
\be
\dd{s^2_{10}}= \left(1+\frac{e^{-4A_0}}{\vo^{2/3}}\right)^{-1/2} \dd{s^2_4+}\left(1+\frac{e^{-4A_0}}{\vo^{2/3}}\right)^{1/2}\dd{s_{CY}^2} \label{eq:warpedmetric3}\:.
\ee
The factor $\left(1+\frac{e^{-4A_0}}{\vo^{2/3}}\right)^{-1/4} \equiv \Omega$ is the redshift factor. In a highly warped region, $e^{-4A_0}\gg \vo^{2/3}$ and $\Omega\sim e^{A_0}\vo^{1/6}\ll 1$.
It is important to keep the following in mind:
   
\begin{enumerate}
\item In highly warped regions  generated  where the local geometry of the underlying CY is close to that of
a conifold, the spacetime metric is close to the KS geometry:
\be
\dd{s^2_{10}}= e^{2A(r)} \dd{s^2_4} + e^{-2A(r)}\left( \dd{r^2} + r^2 \dd{s^2_{T^{1,1}}}  \right)  \label{eq:warpedmetric4} \:,
\ee
The presence of the fluxes resolves the conifold  singularity, and one has a minimal area 3-sphere at the bottom of the throat.
The warp factor takes its minimal value on this 3-sphere \cite{Giddings:2001yu}
\be
\label{amin}
e^{4A_{\rm min}} \sim   e^{-\frac{8\pi K}{3g_sM}} \equiv y \:,
\ee
where $g_s$ is the string coupling and $K$ and $M$ are the integral fluxes that thread the 3-sphere and its dual cycle. 
\item The hierarchy in (\ref{amin}) is related to the value at which the (shrinking) conifold modulus $|z|$ is stabilised \cite{Giddings:2001yu}.
They are related as follows:
$$
y = |z|^{4/3}.
$$
Note that this relation implies that the statistical distribution of the stabilised value of $|z|$ determines the distribution of the hierarchy $y$.

\item The warped volume $\vo_W$ which relates the 10D and 4D Planck masses is given by
\be
\vo_W=\int \dd[6]{y} \sqrt{g_{\small{CY}}} \, e^{-4A} = 
\vo\int \dd[6]{y} \sqrt{g_{CY_0}}\left(1+\frac{e^{-4A_0}}{\vo^{2/3}}\right)\sim \vo\,,
\ee
where the last approximation is valid if the volume of the throat region is small compared to the (large) CY volume.  Note
that the warped volume remains finite even when regions of large warping are present. This is related to the finiteness
of the K\"ahler potential in the complex structure moduli space.

\item The 10D action of an anti-D3 brane can be used to compute its contribution to the 4D scalar potential. This crucially depends on the anti-D3 position $\vec{r}_{{\rm D3}}$ in the internal dimensions, i.e. whether it is in a warped or unwarped region:
\bea
V_{\rm \overline{D3}} = 2T_3\int \dd[4]{x} \sqrt{-g_4}\sim \frac{2M_s^4\,\vo^{2/3}}{e^{-4A(r_{{\rm D3}})}+\vo^{2/3}}&\sim &  \left\{ 
\begin{array}{lcl}
\frac{e^{4A(r_{{\rm D3}})}}{\vo^{4/3}} & \rm{for} & e^{-4A(r_{{\rm D3}})}\gg \vo^{2/3} \\
\frac{1}{\vo^2} & \rm{for} & \vo^{2/3}\gg e^{-4A(r_{{\rm D3}})} \\
\end{array} \right.
\label{eq:uplift}
\eea

where $T_3$ is the tension of the brane. 

\item Note that in the absence of warping one has a string scale contribution to the potential which typically will lead to a run away. Thus warped throats are necessary to obtain stable vacua in the presence of anti-D3 branes. On the other hand, a large volume is necessary to keep the $\alpha'$ expansion valid. In the presence of both large warping and large volume it is important to understand the interplay between these two large quantities and the regime of validity of the effective field theory. This has been analysed in detail in \cite{Giddings:2005ff, Burgess:2006mn, Cicoli:2012fh}. One finds the requirement
\be
e^{-A_0}\ll \vo^{2/3}\ll e^{-4A_0} \,.
\ee
We emphasise that  in the very large radius limit, i.e. $\vo^{2/3}\gg e^{-4A_0(y)}$ at all points in the compact directions, the metric becomes the standard unwarped CY metric $\dd{s_{10}^2}=\dd{s_4^2}+\vo^{1/3}\dd{s_{CY_0}^2}=\dd{s_4^2}+\dd{s_{CY}^2}$.
 In this limit there are no throats, and hence this region of moduli space is not appropriate if one wants to uplift KKLT and LVS AdS vacua.
\end{enumerate}

\subsection{Distributions of $W_0$, string coupling and hierarchy of throats}
\label{DRS}

As emphasised in the introduction, the properties of a string vacuum are determined by the values
 at which moduli are stabilised. Our focus will be on KKLT and LVS models. Here fluxes
 stabilise the complex structure moduli. The K\"ahler moduli continue to remain flat after the
 introduction of fluxes. The stabilisation of the K\"ahler moduli can be studied in a low energy effective
 theory where the complex structure moduli are integrated out. Although the number of fluxes can
 be large, their effect on the low energy effective field theory of the K\"ahler moduli is encoded in terms
 of a small number of parameters. For the observables that  interest us, these are: the expectation value
 of the Gukov-Vafa-Witten superpotential $W_0$, the dilaton $g_s$ and the hierarchy associated with
 the bottom of the warped throat $y$. The statistical distributions of these quantities will serve as input for determining the distributions
 of the observables. The  distributions of $W_0$, $g_s$ and $y$ have been well studied. Below, we describe them.
 \begin{itemize}
  \item The expectation value of the Gukov-Vafa-Witten superpotential $W_0$ is uniformly distributed as a complex
 variable \cite{Denef:2004ze} and physical quantities will be functions of $|W_0|$. Given the flat distribution of $W_0$, the distribution
 for $|W_0|$ is proportional to $|W_0|$.   The distribution of $W_0$ and its physical implications were analysed in detail in \cite{Cicoli:2013swa}. 

 \item The distribution of the dilaton is known to be uniform  \cite{Ashok:2003gk, Denef:2004ze}. In terms of the 
 axio-dilaton $S = s - iC_0$ with $s =1/g_s$
 \be
    \dd{\mathcal{N}} = {\dd{S} \over s^{2}} \, ,
 \ee
 where $\dd{S} = \dd{s}\dd{C_0}$ is the integration measure over the axio-dilaton. This has been confirmed by various numerical and analytic
 studies (see  e.g. \cite{Broeckel:2020fdz, Blanco-Pillado:2020wjn}).

\item Finally, we turn to the hierarchy $y$. As discussed in Sec. \ref{rwt}, the hierarchy is determined by the vacuum expectation
 value of the shrinking conifold modulus $|z|$.  The distribution for $|z|$ (as determined by stabilisation from fluxes) was studied
 in \cite{Denef:2004ze}.  The fraction of vacua in which the conifold modulus takes value below $|z|$ was found to be
 \be
 \label{cfrac}
   f(|z|) =- {C \over \ln{|z|} }\, , 
 \ee
 where $C$ is a positive constant. The corresponding density is
 \be
 \label{cden}
   \mathcal{N}(|z|) \dd{|z|} = {C \dd{|z|}\over |z|(\ln{|z|)^2} }\, .
 \ee
It is important to keep in mind that the singularity in the density for small $|z|$ is benign. Arbitrarily small $|z|$ corresponds to arbitrary large fluxes, which would be in conflict with the D3-tadpole cancellation condition. The statistical description is expected to break down before the singularity. Related is the fact that the fraction of states as given in (\ref{cfrac}) vanishes in the limit of $|z| \to 0$. Although, note that there is a significant enhancement of states in comparison with the expectation from the  canonical metric of $\mathbb{C}$.
 
This distribution was used to study the distribution of throat hierarchies and the expectations for the number of throats in \cite{Hebecker:2006bn}. It was found that throats are ubiquitous.
\end{itemize}
 
Before closing this subsection, we would like to record an important point in the analysis of \cite{Hebecker:2006bn} which will be useful for our study.
 It was argued in \cite{Hebecker:2006bn} that if one is dealing with a CY with a large number of flux quanta, the
 joint distribution of 2 (or a small number of) quantities which are not related through a functional relation
is proportional to the product of the 2 individual distributions. For example, if one considers 2 conifold moduli
$|z_1|$ and $|z_2|$, then these are determined by independent fluxes (hence are functionally independent), and so the joint distribution is
 \begin{equation}
 \label{jtt}
   \mathcal{N}(|z_1|,|z_2|) \dd{|z_1|} \dd{|z_2|} \propto   \mathcal{N}(|z_1|) \mathcal{N} (|z_2|) \dd{|z_1|} \dd{|z_2|} \, .
 \end{equation}
 Similar considerations also apply when one is considering the joint distribution of a conifold modulus and
 a quantity that is a function of a large number of fluxes.
 The product structure in the joint distribution essentially follows from the fact that when there is a large number of fluxes, fixing one quantity should not
 affect the distribution of another quantity significantly unless there is a functional relation between them. See \cite{Hebecker:2006bn} for
 a more detailed discussion.
 
\subsection{Effective field theory of the nilpotent goldstino}
\label{Sec:NilpGold}

    Spontaneous supersymmetry breaking in  (effective) supergravity theories leads to the super-Higgs effect --
 the gravitino eats the goldstino to become massive. If this phenomenon takes place at  energies which are low  compared with the Planck mass, the goldstino couplings can be described by
making use of  a (constrained) independent superfield. Supersymmetry is then non-linearly realised  as in the  Volkov-Akulov formalism. 
 There are  several approaches to describe the low energy dynamics of the goldstino in terms of spurion or constrained superfields (see for instance \cite{Komargodski:2009rz} and references therein). 
Our focus will be on the approach where the goldstino is described in terms of a chiral superfield $X$ that is constrained to be nilpotent, i.e. $X^2=0$. This has the ingredients necessary to describe
supersymmetry breaking  induced by the presence of an anti-D3 brane at the tip of a warped throat in flux compactifications \cite{Bergshoeff:2015jxa,Kallosh:2014wsa,Kallosh:2015nia}.

The effective field theory involving a nilpotent chiral superfield  $X$ can be described in terms of  a K\"ahler potential  $K$, a superpotential $W$ and a gauge kinetic function $f$ whose general forms
are:
\be\label{EFTXVA}
K=K_0  + K_1X + \bar{K}_1\bar{X}\ + K_2X\bar{X} , \qquad W=\rho X + \tilde{W} , \qquad f=f_0+f_1 X,
\ee
where $K_0$, $K_1$, $K_2$, $\rho$, $\tilde{W}$, $f_0$, $f_1$  are functions of other low energy fields. Higher powers of $X$ are absent in $K$ and $W$ since $X^2=0$. The nilpotency condition implies a constraint on the components  of $X$. Expanding $X$ in superspace
\be
X=X_0(y)+\sqrt{2}\psi(y) \theta + F(y)\theta\bar{\theta} \:,
\ee
(where as usual, $y^\mu=x^\mu+ i \theta \sigma^\mu \bar{\theta}$), $X^2=0$  implies 
\be
X_0=\frac{\psi\psi}{2F}\:.
\ee
Thus unless the fermion $\psi$ condenses in the vacuum, the vacuum expectation value of $X_0$ (the scalar component of $X$) vanishes.

For an anti-D3 brane at the bottom of a warped throat of a type IIB flux compactification, the description in  terms of $X$ is very convenient. It allows to treat its effects in terms of supergravity: its contribution to the scalar potential, the gravitino mass and the couplings to moduli and matter fields are easily obtained.  It was shown in \cite{Kallosh:2015nia} that  nilpotent superfield(s) capture all degrees of freedom of an anti-D3 brane  when it  is placed on top of an orientifold plane.  The presence of the fluxes and the orientifold projection leave  the massless goldstino as a low energy propagating degree of freedom, in keeping with the use of a nilpotent superfield $X$ to describe the system. The simplest example is that of an O3-plane and an anti-D3 brane  at the bottom of a warped throat.
In this case, there is no modulus associated with the position of the anti-D3 brane (in contrast to the case of D3-branes in the bulk). This corresponds to the fact that the scalar component of $X$ is not a propagating field. Furthermore,  there is no contribution from $X_0$ to the scalar potential and it can be  consistently set to zero while looking for vacuum solutions.

 Next, let us turn to the form of the K\"ahler potential and superpotential  in (\ref{EFTXVA}).
In the  case of a single K\"ahler modulus $T$ ($T = \vo^{2/3} + i \psi \equiv \tau + i \psi$  is a complex field obtained by pairing the volume modulus and its axionic partner), the functions $\tilde{W}$ and $\rho$ have no dependence on $T$ at the perturbative level, as a result of  holomorphy and the Peccei-Quinn shift symmetry $T\mapsto T+ic$. 
 The zeroth order term in the K\"ahler potential of  \eqref{EFTXVA}, $K_0=-2\ln\vo$ is  invariant (up to a K\"ahler transformation) under the full modular transformation $T\rightarrow (aT-ib)/(icT+d)$ (which is a generalisation of the shift symmetry). If  $X$ transforms suitably, i.e. as a modular form of weight $\kappa$, the quadratic coefficient takes the form $K_2=\beta \tau^{-\kappa}$ (where  $\beta$
 is a constant). Moreover, if the term linear in  $X$ is absent, the only contribution of $X$ to the F-term  potential is a positive definite term:
\be
V_{\rm up}= e^KK^{-1}_{X\bar{X}}\left\| \frac{\partial W}{\partial X}\right\|^2=\frac{|\rho|^2}{\tau^{3-\kappa}}\:.
\ee
This precisely coincides with form of the contribution of an anti-D3 brane at the tip of a warped throat  with hierarchy ${|\rho|^{2} \big{/} \beta}$ as computed from direct dimensional reduction
(see equation \pref{eq:uplift}), if the modular weight is $\kappa=1$\footnote{If $\kappa=0$, one has a $(T + \bar{T})^{-3}$ dependence which corresponds to an  anti-D3-brane in an unwarped region. The magnitude of the  term is of order the string scale $V_{\rm up} \sim M_s^4$, which if included in the low energy theory, would lead to a runaway potential.}.  Since the anti-D3 brane is localised at a particular point in the compactification manifold, direct couplings to gauge fields located at distant D3 or D7-branes are difficult. This implies the
absence of terms linear in $X$ in the gauge kinetic functions. In summary, at leading order in the $\alpha'$ expansion the effective field theory is specified by:
\be
\label{keefftt}
K=-2\ln \vo + \beta \frac{X\bar{X}}{\tau}, \qquad W=\tilde{W} +\rho X, \qquad f=f_0\,,
\ee
where $c$, $\rho$, $\tilde{W}$, $f_0$ are constants and $|\rho|^2/\beta$ is identified with the hierarchy $y$ as defined in \pref{amin}, i.e. $y=|\rho|^2/\beta$.  Furthermore, the superpotential only receives non-perturbative corrections. 
Nilpotency of $X$ implies that the  K\"ahler potential  in  \pref{keefftt} can  be written as:
  \be\label{logKXXb}
K=-3\ln\left(\tau -\frac{\beta}{3} X\bar{X}\right)\:.
\ee
Note that in the regime where the effective field theory is valid, i.e. when the anti-D3 brane is at the tip of a warped throat, the $X$ superfield couples to $T$ in the K\"ahler potential in the same way a superfield $\phi$ describing the D3-brane matter fields,\footnote{Here and in the following we will take a simplified model where we write down only 1 of the 3 complex superfields describing the D3-brane positions. Adding the other 2 would only complicate the expressions without altering our results.} i.e. \cite{Grana:2003ek,Grimm:2004uq}
\be
K_{\text{D3}}=-3\ln\left(\tau -\frac{\alpha}{3}\phi\bar{\phi}\right) \sim -2 \ln \vo + \alpha \frac{\phi\bar{\phi}}{\tau} + \cdots \:,
\ee
where the dots indicate terms which are higher order in the 
 $1/\tau$ expansion. It is then natural to conjecture \cite{Aparicio:2015psl} that the only effect of $X$ in the K\"ahler potential is to shift the K\"ahler coordinate $T$ in the same way as the field $\phi$ does. This was called the {\it log hypothesis}, as it leads one to write the $X\bar{X}$ term inside the log as in \eqref{logKXXb}.

When  both D3-branes and  anti-D3-brane are present, generically the K\"ahler potential can be written as
\be\label{KparamEFT}
K=-2\ln \vo +\alpha\frac{\phi\bar{\phi}}{\tau^\mu}+\beta\frac{X\bar{X}}{\tau^\kappa}+\gamma\frac{X\bar{X}\,\phi\bar{\phi}}{\tau^{\zeta}}+\cdots  \, ,
\ee
with modular weights $\mu=\kappa=1$ as discussed above. Moreover, if $\phi$ and $X$ have modular weights $\mu$ and $\kappa$ respectively, the modular weight for the $X\bar{X}\,\phi\bar{\phi}$ term should be
$\zeta=\mu+\kappa$. In this case, $\zeta=1+1=2$. 
This agrees with the {\it log hypothesis}, i.e. a  K\"ahler potential of the form
\be
K_{\rm no-scale}=-3\ln\left(\tau -\frac{\alpha}{3}\phi\bar{\phi}-\frac{\beta}{3}X\bar{X}\right) \, .
\label{Klog} 
\ee
In fact, expanding this in powers of $1/\tau$, one obtains \eqref{KparamEFT} with the  condition that $\gamma=\frac{\alpha\beta}{3}$. 
The K\"ahler potential in  (\ref{Klog}) is of the standard no-scale form \cite{Cremmer:1983bf, Burgess:2020qsc}. 

  We conclude this subsection with  comments relevant for the computation of soft masses.
In the KKLT scenario, the low energy effective theory is usually written in terms of the  fields with masses of order or below the gravitino mass. These include massless chiral fields (arising as excitations
of open strings) and the K\"ahler moduli. 
Supersymmetry is broken at the minimum of the scalar potential. Both the F-term of $X$ and the F-term of $T$ are different from zero (with $F^T\ll F^X$). Thus, the goldstino is  a combination of the fermion in $X$ and the fermion in $T$.
In LVS, even in the absence of the anti-brane, the overall volume modulus $T_b$ breaks supersymmetry   ($F^{T_b}\neq 0$). Inclusion of  the nilpotent  superfield in the effective action allows one to consider the breaking of supersymmetry induced by fluxes and supersymmetry breaking by the anti-brane on equal footing. Again, the goldstino is a combination of the fermion components of $X$ and of the moduli. Although the dominant component is usually the one from the $T_b$-field, for sequestered models the $X$ component is relevant and its contribution to the soft terms must be accounted for.

\section{Joint gravitino mass and cosmological constant distribution}
\label{sec:3joint}

In this section we will evaluate the joint distribution of the gravitino mass and the cosmological constant for KKLT and LVS models with anti-brane uplifting. As described earlier,
we will make use of the distributions of the axio-dilaton, $W_0$ and $y$ (the hierarchy) to evaluate these. We start by writing the 
expression for the number of vacua in an infinitesimal volume in these coordinates. Making use of the results discussed in Sec. \ref{DRS}, we have
\begin{equation}
\label{distri1}
  \dd{\mathcal{N}} =  {  \eta |W_0|  \over { y (\ln y)^2 s^{2}} }\dd{|W_0|} \dd{S} \dd{y},
\end{equation}
where we have made use of the fact that the hierarchy is related to the size of the shrinking conifold modulus by
$y = |z|^{4/3}$ ($\dd{S}$ is the measure for integration over the axio-dilaton: $\dd{S} = \dd{s}\dd{C_0}$). We have also taken the number of fluxes to be large, so justifying the  factorised form of the density. In both KKLT and LVS the 
cosmological constant and the gravitino mass can be expressed in terms of $|W_0|$, $y$ and $s$. We will use these expressions to carry out a change of variables in \pref{distri1}. This will provide us with the required densities.

\subsection{KKLT}

With the complex structure moduli and the dilaton stabilised by fluxes, we will take the low energy fields to be the K\"ahler moduli, chiral matter and the nilpotent superfield. For simplicity, we will consider one K\"ahler modulus $T$ and matter fields living on D3-branes which we will collectively denote as $\phi$. Then
\begin{equation}
 K = -2\ln{\mathcal{V}} + \tilde{K}_i\ \phi \bar{\phi} + \tilde{Z}_i\ X \bar{X} + \tilde{H}_i\ \phi \bar{\phi} \ X \bar{X} + ... \: ,
\label{paramK\"ahler}
 \end{equation}
where  $\tilde{K}_i$ and $\tilde{Z}_i$ are the K\"ahler metrics for the matter field on D3-branes and the nilpotent goldstino. $\tilde{H}_i $ is the quartic interaction between the matter field and the nilpotent goldstino. As per the  discussion in Sec. \ref{Sec:NilpGold}, these are
\begin{equation}
 \tilde{K}_i  = \frac{\alpha}{\tau}\:, \qquad \tilde{Z}_i  = \frac{\beta}{\tau}\:,\qquad \tilde{H}_i  = \frac{\gamma}{\tau^{2}}\:,
\label{mattermetric}
 \end{equation}
where the scaling of $\tilde{K}_i$ with $\tau$ is due to the modular weight of the matter field on D3-branes \cite{Conlon:2006tj,Aparicio:2008wh}. The superpotential is
\begin{equation}
 W \ =\ W_{0} +\rho X  + A\ e^{-a T}\:,
\label{Wnil}
 \end{equation}
 where we have included a non-perturbative contribution in $T$ which is needed to stabilise the K\"ahler modulus. The KKLT construction requires $|W_0| \ll 1$ (see e.g. \cite{Demirtas:2021nlu, Demirtas:2019sip, Cole:2019enn, Broeckel:2021uty, Cicoli:2022vny} and references therein for recent works on explicit construction of vacua with low values of $|W_0|$).
 We will study the statistical distributions for
 fixed values of $a$ and $A$.

Recall that the supergravity scalar potential is determined in terms of the K\"ahler potential and the superpotential as
 \begin{equation}\label{sugrapotential}
  V = e^{K} \left(  K^{I\bar{J}} D_I W D_{\bar{J}}\bar{W} - 3 |W|^2  \right) \qquad \mbox{with} \qquad D_IW = \partial_I W + K_I W \: ,
 \end{equation}
where the indices  $I$ and $J$ run over the chiral superfields $T,\phi,X$.  $K^{I\bar{J}}$ is the inverse of the matrix $K_{I\bar{J}}\equiv \partial_I\partial_{\bar{J}} K$ and $K_I\equiv \partial_IK$. It
is often convenient to write \pref{sugrapotential} as
\begin{equation}\label{sugrapotential2}
V =  F^{I} F_I  - 3 m_{3/2}^2 \, ,
\end{equation}
where $F_I \equiv e^{K/2} D_IW$ and $F^I \equiv e^{K/2}K^{I\bar{J}} D_{\bar{J}}\bar{W}$ are the F-terms. The gravitino mass is instead given by $m_{3/2}\equiv e^{K/2}|W|$.

By making use of  \eqref{paramK\"ahler} and \eqref{Wnil} in \eqref{sugrapotential},  one finds
\begin{equation}
\label{ScPotKKLTUp}
 V = \left(V_{\text{KKLT}} + V_{\text{up}}\right) + \frac{2}{3}\left(  (V_\text{KKLT}  +   V_{\text{up}}) + \frac{1}{2} V_{\text{up}} \left(1-\frac{3 \gamma}{\alpha \beta} \right)  \right) \ |\hat{\phi}|^2\:,
\end{equation}
where  the scalar component of the nilpotent field is set to zero (assuming no condensation of fermions). 
 $V_\text{KKLT}$ is the  KKLT potential in the absence of the uplifting term:
\begin{equation}
 V_{\text{KKLT}} \ =\ \frac{2\ e^{-2 a \tau} a A^2}{s\ \mathcal{V}^{4/3}} + \frac{2\ e^{-2 a \tau} a^2 A^2}{3s\ \mathcal{V}^{2/3}}\ - \frac{2\ e^{- a \tau} a A\ W_0}{s\ \mathcal{V}^{4/3}}\,.
\label{kklt3} 
 \end{equation}
To reduce clutter in the equations we have followed the standard practice of writing both $W_0$ and $A$ real and positive.  For general values of $W_0$ and $A$ the expression for the potential can be obtained from the above by taking $W_0 \to |W_0|$ and $A \to |A|$. We have also neglected the overall contribution from the K\"ahler potential for the complex structure moduli which is
an order one multiplicative factor. The imaginary part of $T$ (the axion $\psi$) has its minimum at $\psi = \pi/a$. This  is responsible for the negative sign in the third term in  (\ref{kklt3}). The uplift term $V_{\text{up}}$ arises from $F^X F_X$ and is given by
\begin{equation}
\label{VupKKLT}
 V_{\text{up}} = \frac{\rho^2}{2\beta s \tau^{2}} \equiv \frac{y}{2 s \tau^2}\qquad\text{with}\qquad y = \frac{\rho^2}{\beta}\,.
\end{equation}
The third contribution in (\ref{ScPotKKLTUp}) is proportional to $|\hat{\phi}|^2$ with $\hat\phi$ denoting the canonically normalised matter scalar field $\phi$. This contribution corresponds to its soft mass. For non-tachyonic scalars, the vacuum expectation value of $\hat\phi$ would vanish, and so we will proceed by setting $\hat{\phi} =0$. Soft masses will be discussed in Sec. \ref{sec:4disc}.\footnote{If the log hypothesis \pref{Klog} holds, at leading order in the $\alpha'$ expansion, the soft masses for D3-brane matter vanish in KKLT. $\alpha'$ corrections and anomaly mediation contributions are relevant. $\alpha'$ corrections always make a positive contribution to the square masses.}

Minimising the scalar potential, one finds that the following holds at the minimum
\begin{equation}
  \label{kkltmin4}
    W_0 = e^{-a \tau} A \left(1 + \frac{2}{3} a \tau + \frac{y\,e^{2 a \tau}}{2 a^2 A^2 \tau}\right) \,,
\end{equation}
where we have dropped subleading terms in the $(a\tau)^{-1}$ expansion multiplying the term proportional to $y$. In what follows, we shall drop such subleading
terms, but shall include their effect in the numerical results that we will present later.
 Using   \eqref{kkltmin4} in the expression for the scalar potential \eqref{ScPotKKLTUp}, its value at the minimum is found to be
\begin{equation}
\label{ScPotKKLTtot}
 \Lambda = - \frac{2 e^{-2 a \tau} a^2 A^2}{3s\ \tau} + \frac{y}{2 s \tau^{2}} \equiv \left(V_{\text{KKLT}}^{(0)} + V_{\text{up}}\right).
\end{equation}
Note that the hierarchy $y$ can be tuned to make the cosmological constant zero or extremely small and positive. For a Minkowski vacuum one needs
 \begin{equation}
\label{dScondRho}
  y= \frac{4 }{3}\, \tau \, e^{-2a\tau} a^2 A^2 \,.
 \end{equation}
The gravitino mass becomes
\begin{equation}
\label{m32kklt}
m_{3/2} = e^{K/2}|W| = { 1 \over \sqrt{2s} } { 1 \over \tau^{3/2} } \left( W_ 0 - A\, e^{-a \tau} \right).
\end{equation}
Next, we turn to evaluating the joint distribution for the gravitino and cosmological constant. For this, we will perform the change of variables $ (W_0, y, S) \to (m_{3/2}, \Lambda, S)$ in \pref{distri1}. We need to compute the Jacobian of this transformation. The expressions for
the cosmological constant and the gravitino mass (\pref{ScPotKKLTtot} and \pref{m32kklt}) have explicit dependence on 
$(W_0, y, s)$ and also implicit dependence via  $\tau$. To compute the partial derivatives of the cosmological constant and the gravitino mass
with respect to $W_0$ and $y$ we need to compute the partial derivatives of $\tau$ with respect to these variables. Making use of
\pref{kkltmin4}, we find
\begin{equation}
\label{twopar}
{\del \tau \over \del y}  = - { e^{ a \tau} \over 2 A a^{2} \tau f(\tau) } \, ,
\end{equation}
with 
\begin{equation}
\label{fdef}
 f(\tau) = A e^{-a \tau} \left( -{2 \over 3} a^{2} \tau \left( 1  + {1 \over 2 a \tau} \right)   +
   { {y \,e^{2 a \tau}} \over {2 a A^{2} \tau}} \left( 1 - {1 \over a \tau }\right) \right) \, .
\end{equation}
Now let us turn to the entries of the Jacobian. These are
\begin{equation}
\label{mw}
{\del m_{3/2} \over \del W_0 } =   { 1 \over \sqrt{2s} } { 1 \over \tau^{3/2} } +
\left( { 1 \over \sqrt{2s} } { 1 \over \tau^{3/2} }  a A e^{-a \tau} - {3 \over 2 \tau} m_{3/2}   \right) { \del \tau \over \del W_0 } \, ,
\end{equation}
\begin{equation}
\label{MW1}
{\del m_{3/2} \over \del y } =  
\left( { 1 \over \sqrt{2s} } { 1 \over \tau^{3/2} }  a A e^{-a \tau} - {3 \over 2 \tau} m_{3/2}   \right) { \del \tau \over \del y } \, .
\end{equation}
We also have
\begin{equation}
\label{ljac}
 {\del \Lambda \over \del W_0} = { \del \Lambda \over \del \tau} { \del \tau \over \del W_0}\, , \qquad\qquad
  {\del \Lambda \over \del y} = { \del \Lambda \over \del \tau} { \del \tau \over \del y} + {1 \over 2 s \tau^{2} } \, ,
\end{equation}
 with 
\begin{equation}
\label{ltau}
{ \del \Lambda \over \del \tau } = { 4 e^{-2 a \tau} a^{3} A^{2} \over 3 s \tau} \left( 1 + {1 \over 2 a \tau} \right)
      - { y \over s \tau^{3} } \, ,
\end{equation}

\subsubsection{Analytical estimate}

The expressions in (\ref{mw}), (\ref{MW1}) and (\ref{ljac}) are rather cumbersome. While we will use them for
our numerical analysis in Sec. \ref{Numerics}, we continue our analytic investigation in a regime
that leads to considerable simplifications. For this, we define the quantity $x$ as
 \begin{equation}
 \label{xdef}
 x \equiv  { {y\, e^{2 a \tau}} \over {2 a^{2} A^{2} \tau}}\, ,
 \end{equation}
which allows to write (\ref{ScPotKKLTtot}) as
\begin{equation}
 \Lambda = - \frac{2 e^{-2 a \tau} a^2 A^2}{3s\,\tau} \left(1-\frac32\,x \right) = -3 m_{3/2}^2 \left(1-\frac32\,x\right)
 \left(1+\frac{3}{2 a \tau}\,x\right)^{-2}\,.
 \label{LambdaKKLT}
\end{equation}
We shall now focus on the regime in which $y$ is varied such that $0 < x \lesssim \mathcal{O}(1) \ll (a\tau)$. Comparing with (\ref{LambdaKKLT}) we see that this allows for uplift to zero and positive values of the cosmological constant. Hence, {\it {the regime covers the region of most interest}} since $x \gg 1$ corresponds to an unstable situation where the uplifting contribution would yield a runaway. Also note that in this regime, the contribution of the term involving  $y$ in right hand side of (\ref{kkltmin4}) becomes subdominant. We can therefore take the approximation
 \begin{equation}
 \label{wsim}
   W_0  = e^{-a \tau} A \left(1 + \frac{2}{3} a \tau + x\right) \simeq {2 \over 3} A (a\tau) e^{-a \tau} \,,
 \end{equation}
 and so, to leading order in the $(a \tau)^{-1}$ expansion, we have
 \begin{equation}
 \label{tsol}
 \tau \simeq -{ 1 \over a} \ln \left( {3 \over 2A} W_0 \right),
 \end{equation}
and the gravitino mass looks like
\begin{equation}
\label{mtt}
 m_{3/2} \simeq {W_0 \over { \sqrt{2s} \tau^{3/2} }} = {2 \over 3} { { (a\tau) A} \over { \sqrt{2s} \tau^{3/2} } } e^{- a \tau}\,.
 \end{equation}
This allows us to write
\begin{equation}
\label{TL}
\tau \simeq - {1 \over a} \ln \left(  3 \sqrt{ {s  \over 2} } { {m_{3/2} \over {a^{3/2} A} }}  \right)\, .
 \end{equation}
In the regime we are considering, we can write \pref{ScPotKKLTtot} as
\begin{equation}
\label{yl}
y \simeq 2 s \tau^{2} \left(  \Lambda + 3 m^2_{3/2}  \right)\, .
\end{equation}
 Combining this with (\ref{TL}) we can express $y$ in terms of $\Lambda$, $m_{3/2}$ and $s$. Similarly, making use
 of  (\ref{TL}) in \pref{mtt} gives $W_0$ in terms of the same quantities.
  
The Jacobian entries undergo significant simplifications in this regime where  the function $f(\tau)$ defined in (\ref{fdef}) becomes $f(\tau) \simeq  -{2 \over 3} a^{2} A e^{-a \tau}  \tau$.
We then have
  \begin{equation}
  \label{tp}
  { \del \tau \over \del W_0} \simeq - {3 \over 2 A a^{2} \tau} e^{a \tau} \,,\qquad\qquad  { \del \tau \over \del y} \simeq {3 \over 4 A^2 a^{4} \tau^2} e^{2a \tau}\, .
    \end{equation}
Let us turn to the partial derivatives of $m_{3/2}$. First note that by making use of (\ref{kkltmin4}) and (\ref{m32kklt}) one can write
\begin{equation}
\label{m3oth}
m_{3/2} = { A e^{-a \tau} \over \tau^{3/2} (2s)^{1/2}} \left( {2 \over 3} a \tau + { y \,e^{2a\tau} \over 2 a^{2} A^{2} \tau} \right)
\equiv  { A e^{-a \tau} \over \tau^{3/2} (2s)^{1/2}} \left( {2 \over 3} a \tau + x  \right) \, .
\end{equation} 
 Recall now that 
\begin{equation}
\label{mw1}
{\del m_{3/2} \over \del W_0 } =   { 1 \over \sqrt{2s} } { 1 \over \tau^{3/2} } +
\left( { 1 \over \sqrt{2s} } { 1 \over \tau^{3/2} }  a A e^{-a \tau} - {3 \over 2 \tau} m_{3/2}   \right) { \del \tau \over \del W_0 }\,.
\end{equation}
Making use of $\pref{m3oth}$, we see that
\begin{equation}
\label{step}
\left( { 1 \over \sqrt{2s} } { 1 \over \tau^{3/2} }  a A e^{-a \tau} - {3 \over 2 \tau} m_{3/2}   \right) =   - {3 \over 2 }{ A e^{-a \tau} \over \tau^{5/2} (2s)^{1/2}}\, x\,.
\end{equation}
Using (\ref{tp}) and (\ref{step}) in \pref{mw1}, and the fact that in the above equation $0<x\lesssim \mathcal{O}(1)$, we see that the term proportional to $ { \del \tau \over \del W_0 }$ is always subleading in \pref{mw1}. Therefore we have 
\begin{equation}
\label{p1f}
{\del m_{3/2} \over \del W_0 } \simeq  { 1 \over \sqrt{2s} } { 1 \over \tau^{3/2} } \, .
\end{equation}
Similarly, using the above formulae, one finds
\begin{equation}
\label{p2f}
  { \del m_{3/2} \over \del y} \simeq - {9 \sqrt{a} \over 8 A}  { 1 \over \sqrt{2s} }  { 1 \over  (a \tau)^{9/2} }\, e^{a \tau} x \, .
\end{equation}
Now, let us come to the derivatives of $\Lambda$. Note that the ratio of the two terms in \pref{ScPotKKLTtot} is
$-{3 \over 4} x$. Given this (and the fact that $0<x\lesssim \mathcal{O}(1)$), comparing the various terms in $\pref{ltau}$ gives 
\begin{equation}
\label{ltsim}
{\del \Lambda \over \del \tau} \simeq  { 4 e^{-2 a \tau} a^{3} A^{2} \over 3 s \tau} \, .  
\end{equation}
Combining this with \pref{ljac} and \pref{tp} one finds
\begin{equation}
\label{lwo}
{\del \Lambda \over \del W_0} \simeq -{ 2A a^{3} \over s} { e^{-a \tau} \over (a \tau)^2 } \qquad\text{and}\qquad
{\del \Lambda \over \del y}  \simeq {1 \over 2s \tau^{2} } \, .
\end{equation}
Combining (\ref{p1f}), (\ref{p2f}) and (\ref{lwo}), the Jacobian is
\begin{equation}
\label{jacfin}
  J \simeq {1 \over (2s)^{3/2} } {1 \over \tau^{7/2} } \, .
\end{equation}
Finally, we have
\begin{equation}
\label{distriKK}
  \dd{\mathcal{N}} =  {  \eta \,W_0  \over { y (\ln y)^2 s^{2}} }\dd{W_0} \dd{y} \dd{s} = {  \eta \,W_0  \over { y (\ln y)^2 s^{2}} }  |J|^{-1} \dd{m_{3/2}} \dd{\Lambda}  \dd{S} \, .
  \end{equation}
Equations (\ref{mtt}), (\ref{TL}) and (\ref{yl}) can be used to express this density in terms of the desired quantities. We find
\begin{equation}
\label{kkdis}
\dd{\mathcal{N}} = {  2 \tau^{3} m_{3/2} \over { \left( 3 m^2_{3/2} +  \Lambda\right)  \left[ \ln  \left( 2s\tau^{2} (3 m^2_{3/2} +  \Lambda) \right) \right]^{2}s }}\,\dd{m_{3/2}} \dd{\Lambda} \dd{S}  \, .
\end{equation}
To get the expression for the joint distribution of $m_{3/2}$ and $\Lambda$, we need to perform the integration over the axio-dilaton moduli space. However, the important features can be extracted by analysing the above density at fixed values of the axio-dilaton $(s,C_0)$. Note that in the $|\Lambda| \ll m^2_{3/2}$ limit (which is physically most interesting) the density scales as
\begin{equation}
\rho_{\rm KKLT}(\Lambda \simeq 0) \simeq {\ln m_{3/2} \over m_{3/2} }\,,
\label{rhoKKLT}
\end{equation} 
implying that it is tilted favourably towards lower values of $m_{3/2}$.

This result can be understood as follows. Let us think of uplifting various AdS vacua obtained before the introduction of the uplift term.
These vacua are supersymmetric and $\Lambda = - 3 m_{3/2}^2$. The introduction of the uplift term has a very small effect on the value of $m_{3/2}$. Thus, when we consider vacua with cosmological constant close to zero, a vacuum with a low value of $m_{3/2}$ also has  a low value of the hierarchy $y$ (as the associated supersymmetric AdS vacuum before the introduction of the uplift has a $\Lambda$ of small magnitude). The distribution of throats is such that it is tilted in favour of lower values of $y$, and this makes the distribution of $m_{3/2}$ favourable towards lower values of $m_{3/2}$.

\subsubsection{Numerical results}
\label{Numerics}

\begin{figure}[!htbp]
\centering
\includegraphics[scale = 1.3]{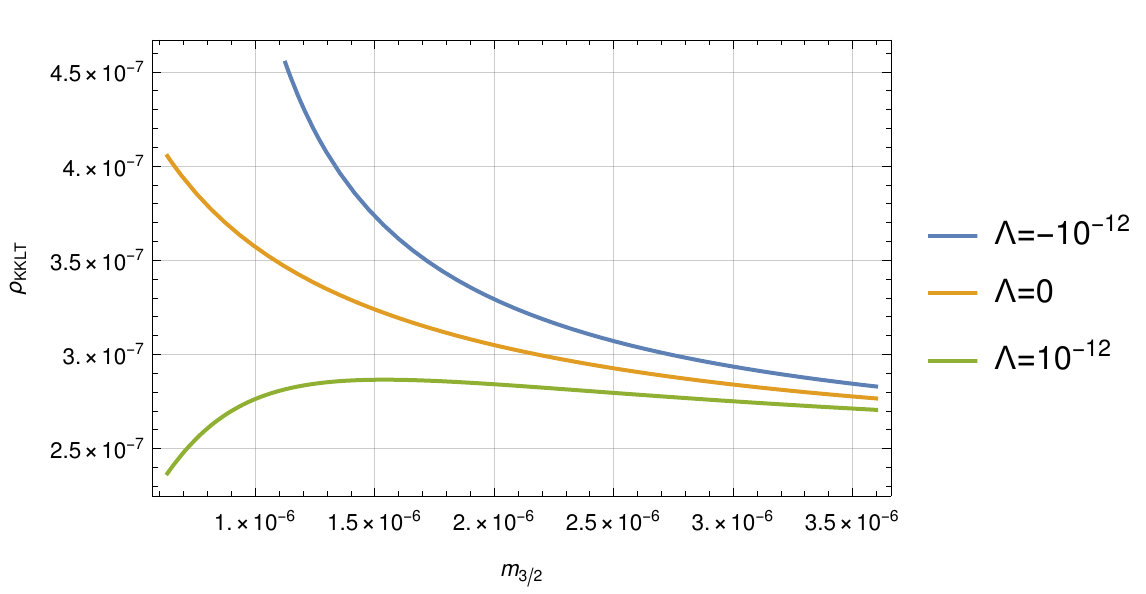}
\caption{Density of KKLT flux vacua at fixed values of the cosmological constant $\Lambda$ as a function of the gravitino mass $m_{3/2}$ in Planck units.}
\label{Fig1}
\end{figure}

Let us now present the numerical results which we have obtained for the joint distribution of the cosmological constant and the gravitino mass without making any approximation. Fig. \ref{Fig1} shows the density of states as a function of $m_{3/2}$ for different fixed values of $\Lambda$ corresponding to AdS, Minkowski and dS vacua. In the physically interesting region with $|\Lambda| \ll m^2_{3/2}$, the 3 curves approach each other, reproducing the analytical estimate (\ref{rhoKKLT}) where $\rho_{\rm KKLT}$ becomes independent of $\Lambda$ and is inversely proportional to $m_{3/2}$ (up to a logarithmic dependence). Note that the green curve with positive $\Lambda$ features a raising behaviour of $\rho_{\rm KKLT}$ for $m_{3/2}^2\lesssim \Lambda$. However this regime can be ignored for the following two reasons: ($i$) it is valid only for a small window of values of $m_{3/2}$ close to $m_{3/2}^2\lesssim \Lambda$ since, as can be seen from (\ref{LambdaKKLT}), $m_{3/2}^2\ll \Lambda$ would require $x\gg 1$ that yields a runaway; ($ii$) the region characterised by $m_{3/2}^2\lesssim \Lambda$ is phenomenologically irrelevant since it would correspond to an essentially massless gravitino with $m_{3/2}$ below the Hubble parameter.

\begin{figure}[!htbp]
\centering
\includegraphics[scale = 1.3]{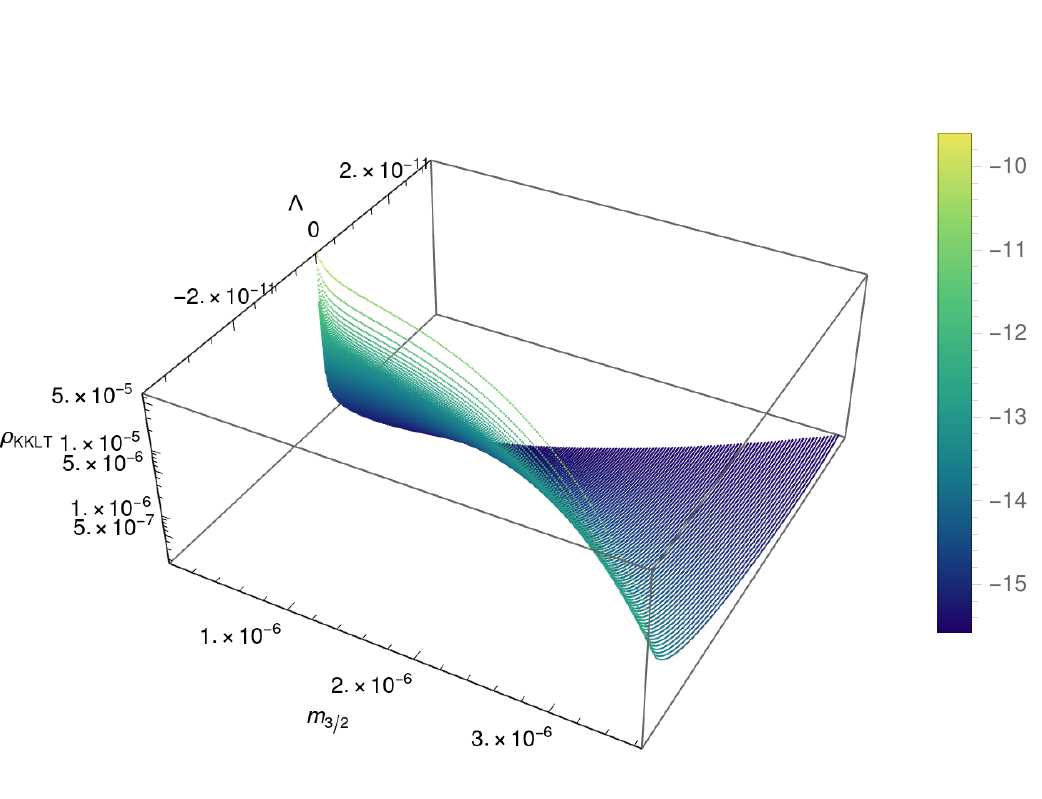}
\caption{Density of KKLT flux vacua as a function of the gravitino mass $m_{3/2}$ and the cosmological constant $\Lambda$ in Planck units.}
\label{Fig2}
\end{figure}

For completeness, in Fig. \ref{Fig2} we have also plotted the density of vacua as a function of $m_{3/2}$ and $\Lambda$. Note that the two blank regions corresponds respectively to the runaway limit (for positive $\Lambda$) and to the violation of the supergravity lower limit $\Lambda \geq - 3 m_{3/2}^2$ (for negative $\Lambda$).

\subsection{LVS}
\label{Sec:LVS}

We now turn to the LVS models. Using the same notation as in the previous subsection, the K\"ahler potential can be written as 
\begin{equation}
K = -2\ln \left(\mathcal{V}+ \xi\,s^{3/2} \right)  
+ \tilde{K}_i\ \phi \bar{\phi} + \tilde{Z}_i\ X \bar{X} + \tilde{H}_i\ \phi \bar{\phi} \ X \bar{X} + ... \:.
\label{K\"ahlerlvs}
\end{equation}
We will focus on the simplest  LVS example, with two  K\"ahler moduli $T_s$ and $T_b$ with real parts  $\tau_b$ and $\tau_s$, with the CY volume  having a Swiss cheese structure: $\mathcal{V} = \tau_b^{3/2}-\tau_s^{3/2}$. The coefficients $\tilde{K}_i$, $\tilde{Z}_i$ and $\tilde{H}_i$ are matter metric and quartic interaction coefficients. 
The superpotential is 
\begin{equation}
 W \ =\ W_{0} +\rho X  + A\ e^{-a_s T_s}\,
\end{equation}
where $W_0\sim \mathcal{O}(1-10)$ in LVS. The supergravity scalar potential (after setting the matter field expectation values to zero) takes the  form:
\begin{equation}
  V = V_\text{LVS} + V_{\text{up}} \, ,
  \label{Vtotlvs}
\end{equation}
where $V_\text{LVS}$ is the LVS potential in the absence of the uplift sector
\begin{equation}
V_\text{LVS} = \frac{4}{3}\frac{e^{-2 a_s \tau_s}\sqrt{\tau_s} \ a_s^2 A^2 }{s\ \mathcal{V}}  - \frac{2 e^{- a_s \tau_s} \tau_s \ a_s A\ W_0 }{s\ \mathcal{V}^2} + \frac{3 \sqrt{s}\ \xi\ W_0^2}{8\ \mathcal{V}^3 } \, .
\end{equation}

In LVS the minimum is non-supersymmetric  before the introduction of the uplifting term. Minimising the LVS potential $V_\text{LVS}$, one obtains the  conditions that the K\"ahler moduli have to satisfy in the minimum\footnote{We will work to leading order in the $(a_s\tau_s)^{-1}$ expansion.}:
\begin{equation}
e^{-a_s \tau_s} = \frac{3 \  \tau_s^{1/2} \ W_0}{4 a_s A \ \mathcal{V}} \, ,
\label{minlvs1}
\end{equation}
and 
\begin{equation}
\tau_s^{3/2} = \frac{s^{3/2} \xi}{2}
\label{minlvs2} \, .
\end{equation}
The value of the potential at this minimum is
\begin{equation}
 V_\text{LVS}^{(0)} = -\frac{3 \sqrt{s}\ \xi \ W_0^2}{16 a_s \tau_s \ \mathcal{V}^3}\,.
 \label{CCseq}
\end{equation}
Next, we consider the potential (\ref{Vtotlvs}) that includes the $X$-contribution responsible for uplifting the AdS minimum. The minimum condition  (\ref{minlvs1}) is not modified, while  (\ref{minlvs2}) is changed to
 \begin{equation}
 \label{tsn}
   \tau_s^{3/2} ={s^{3/2} \xi \over 2}\left(1 + {16 \over 27} {  {\mathcal{V}^{5/3} y} \over {W_0^2{s^{3/2} \xi }} } \right) \, .
    \end{equation}
 For later use, we define
 \begin{equation}
 \label{xlvs}
x_{\rm LVS} \equiv {16 \over 27} {   y \,(a_s \tau_s) {\mathcal{V}^{5/3}} \over {W_0^2{s^{3/2} \xi }} } \, .
 \end{equation}
The value of the potential at the minimum now becomes
\begin{equation}
\label{lambvs}
\Lambda = V_\text{LVS}^{(0)} + {5 \over 9}\, V_{\rm up} =  -  { 3\sqrt{s} \xi W_0^{2} \over 16 a_s \tau_s \mathcal{V}^{3} }  + {5 \over 9} {y  \over 2 s \mathcal{V}^{4/3} } = 
-  { 3\sqrt{s} \xi W_0^{2} \over 16 a_s \tau_s \mathcal{V}^{3} } \left(1 -\frac52 \,x_{\rm LVS}\right)\,,
\end{equation}
which gives a Minkowski minimum for
 \begin{equation}
 \label{mcon}
   y = {27 \over 40}  {   {W_0^2{s^{3/2} \xi } }\over {{\mathcal{V}^{5/3} a_s \tau_s}} }\qquad\Leftrightarrow\qquad x_{\rm LVS} = \frac{2}{5}\,.
 \end{equation}
Thus we realise that (\ref{lambvs}) can allow for uplift to zero and positive values of the cosmological constant provided $y$ is varied so that $x_{\rm LVS}$ lies in the $0 < x_{\rm LVS} \lesssim \mathcal{O}(1) \ll (a_s\tau_s)$ regime, with $x_{\rm LVS} \gg 1$ corresponding to the runaway limit. Hence from now on we shall focus just on the interesting region where the potential is stable and the gravitino mass is given by
\begin{equation}
\label{gvi}
  m_{3/2} = { W_0 \over {\sqrt{2s}\, \mathcal{V}}} \, .
\end{equation}

\begin{figure}[!htbp]
\centering
\includegraphics[scale = 1.3]{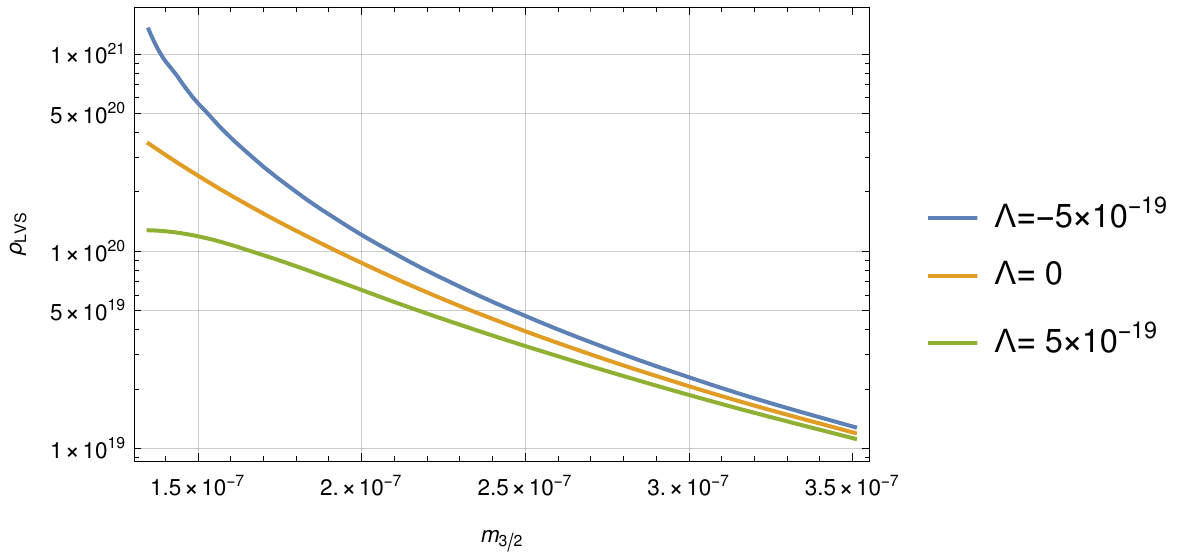}
\caption{Density of LVS flux vacua at fixed values of the cosmological constant $\Lambda$ as a function of the gravitino mass $m_{3/2}$ in Planck units.}
\label{FigThree}
\end{figure}

\begin{figure}[!htbp]
\centering
\includegraphics[scale = 1.3]{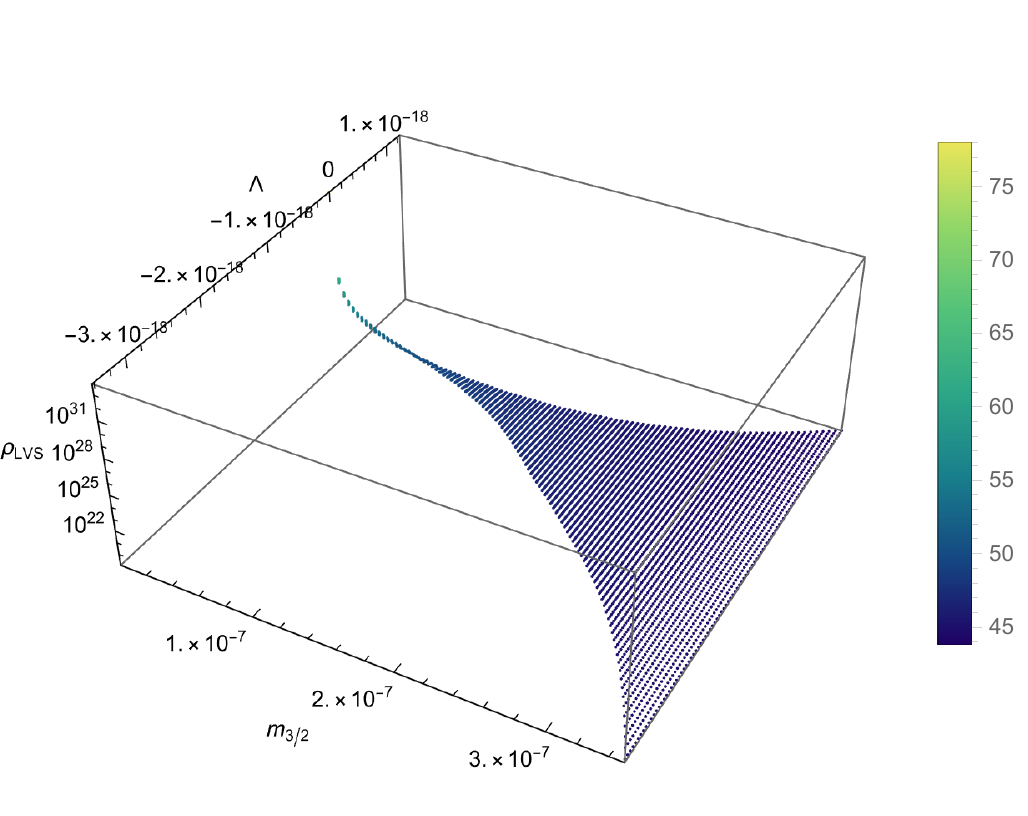}
\caption{Density of LVS flux vacua as a function of the gravitino mass $m_{3/2}$ and the cosmological constant $\Lambda$ in Planck units.}
\label{FigFour}
\end{figure}

Let us now perform a variable change in \pref{distri1} to obtain the joint distribution of the gravitino mass and the cosmological
constant: $ (W_0, y, s, C_0) \to (W_0, \Lambda, m_{3/2}, C_0)$.  $\Lambda$ and $m_{3/2}$, given in (\ref{lambvs}) and (\ref{gvi}), have explicit dependence on $s$ and $y$ and also implicitly depend on the dilaton and the hierarchy via $\tau_s$ and $\cal{V}$. Note that, at leading order in the ${(a_s \tau_s)^{-1} }$ expansion, we have  ${\del \mathcal{V} \over \del \tau_s} =  a_s \mathcal{V}$ and
 \begin{equation}
 \label{tds}
 {\del \tau_s \over \del s} = {  \left( { \xi \over 2} \right)^{2/3} \over { 1- {10 \over 9} x_{\rm LVS}}} \equiv  {  \left( { \xi \over 2} \right)^{2/3} \over \Psi}\,,
\qquad
 {\del \tau_s \over \del y} =  \frac{32}{81}
 {  {\mathcal{V}^{5/3} \tau_s } \over {W_0^2{s^{3/2} \xi }}} { 1 \over \Psi} = \frac23{ {x_{\rm LVS}} \over y\,a_s \,\Psi}\, .
   \end{equation}
  Now we turn to the entries of the Jacobian (again, to leading order in the $(a_s \tau_s)^{-1}$ expansion). They read
  \begin{equation}
\label{grs}
 {\del m_{3/2} \over \del s} = -a_s m_{3/2} { \del \tau_s \over \del s}\,, \qquad  {\del m_{3/2} \over \del y} = -a_s m_{3/2} { \del \tau_s \over \del y}\, ,
 \end{equation}
\begin{eqnarray}
\label{lamsd}
{\del \Lambda \over \del s} &=& -3 V_{\rm LVS}^{(0)} a_s \Psi {\del \tau_{s} \over \del s} = -3 V_{\rm LVS}^{(0)} a_s \left( {\xi \over 2} \right)^{2/3}\,,
 \qquad  {\del \Lambda \over \del y} =  {1\over 2 s \mathcal{V}^{4/3} }\,,
\end{eqnarray}
where we have used \pref{tds}. These give a Jacobian of the form
\begin{equation}
\label{jfin1}
  J =  - {5 \over 18} {{ m_{3/2} a_s \over \Psi}} \left( { \xi \over 2} \right)^{2/3} {1 \over {s \mathcal{V}^{4/3} }}  = - { 5 \over  2^{1/3} \, 9 }\left( { \xi \over 2} \right)^{2/3} { { a_s m_{3/2}^{7/3} \over \Psi s^{1/3} W_0^{4/3}}} \,.
  \end{equation}
From \pref{lambvs} we can write
\begin{equation}
\label{lambvs2}
y =  {18 \over 5} s \mathcal{V} ^{4/3}  \left( \Lambda +    { 3\sqrt{s} \xi W_0^2 \over 16 a_s \tau_s \mathcal{V}^{3} }  \right) =   {2^{1/3}\,9 \over 5} { s^{1/3} W_0^{4/3}   \over m_{3/2}^{4/3} }
\left( \Lambda +    { 3 \sqrt{2}  \xi m_{3/2}^{3} s^{2} \over 8 a_s \tau_s W_0 }  \right) \, .
\end{equation}
In summary, we find
\begin{equation}
\label{distri1l}
  \dd{\mathcal{N}} =  {  \eta \,W_0  \over { y (\ln y)^2 s^{2}} } |J|^{-1} \dd{W_0} \dd{m_{3/2}} \dd{\Lambda} \dd{C_0},
\end{equation}
with $J$ and $y$ given in \pref{jfin1} and \pref{lambvs2} and $\tau_s = s \left( { \xi \over 2} \right)^{2/3} \simeq - {1 \over a_s} \ln m_{3/2}$. To get the final form of the distribution, one should integrate over $C_0$ and $W_0$. Even if this is not tractable, the interesting features of the joint distribution can be 
obtained by considering fixed values of $C_0$ and $W_0$ (which is an $\mathcal{O}(1-10)$ quantity in LVS). In the physically interesting regime where $|\Lambda| \ll m_{3/2}^{3}$, we can also make the following approximations
\begin{equation}
\label{lvsesti}
 |J|^{-1} \simeq { ( \ln m_{3/2})^{1/3} \over m_{3/2}^{7/3}}\,,\qquad y^{-1} \simeq {  1\over (\ln m_{3/2})^{4/3}  m_{3/2}^{5/3}}\, ,
 \end{equation}
which give
\begin{equation}
  \rho_{\rm LVS}(\Lambda \simeq 0) \simeq {1 \over (\ln m_{3/2})^{3}m_{3/2}^{4} } \,.
\end{equation}
Interestingly, we find that the distribution of flux vacua with cosmological constant close to zero is highly tilted towards lower values of $m_{3/2}$. We have confirmed this analytical estimate with an exact numerical evaluation of the density of LVS flux vacua as a function of $m_{3/2}$ and $\Lambda$. Two plots showing these results are presented in Fig. \ref{FigThree} and Fig. \ref{FigFour}. As stressed already in the KKLT case, the blank region in Fig. \ref{FigFour} for $\Lambda >0$ would lead to a runaway, while the blank region for $\Lambda<0$ would correspond to a violation of the supergravity lower limit $\Lambda \geq - 3 m_{3/2}^2$.

\section{Discussion and conclusions}
\label{sec:4disc}

In this paper we have studied the joint distribution of the gravitino mass and cosmological constant in KKLT
and LVS models with the uplift sector arising from an anti-brane at the tip of a warped throat. We have found that, at 
values of the cosmological constant close to zero, the gravitino mass distribution is tilted favourably towards
lower scales. This result is different from that based on generic expectations for the size of F- and D-terms \cite{Douglas:2004qg, Susskind:2004uv}, including also moduli stabilisation \cite{Broeckel:2020fdz}. The form of
the distribution of the throat hierarchies and the nature of the AdS vacua before the introduction of the uplift sector\footnote{This is one of the factors that contributes to the difference in the result for KKLT and LVS models.} leads to this difference. This work gives strong motivation for similar studies in related setups so that a better understanding of the joint statistics of the scale of supersymmetry breaking and the cosmological constant in string vacua can be obtained. 

Our results have several interesting implications which we now briefly discuss:
\begin{itemize}
\item{\bf{Distribution of soft terms:}} Supersymmetry breaking in a hidden sector leads to the generation of soft terms in the visible sector. The strength of
 supersymmetry breaking in the visible sector is characterised by the size of the scalar masses $m_0^2$, gaugino masses $M_{1/2}$ and trilinear couplings $A_{ijk}$. In both KKLT and LVS, these depend on how the Standard Model is realised on either D3- or D7-branes. Soft terms for both realisations were analysed
in  \cite{Aparicio:2015psl} (see also \cite{Aparicio:2014wxa, Aparicio:2015sda, Kallosh:2015nia, Choi:2005ge,Choi:2005uz,Falkowski:2005ck,Lebedev:2006qq}). In the case of D7-branes, all soft terms tend to be of order the gravitino mass. On the other hand, in the case of D3-branes, the models can exhibit sequestering at tree level with non-zero soft masses generated by $\alpha'$ and quantum corrections. One can find interesting patterns in the structure of soft masses, allowing for the realisation of MSSM-like spectra or (mini) split supersymmetry. All the soft parameters are of
the form $m_{3/2}^{1+p}$ with $p \geq 0$ or  $m_{3/2} \big{/} (\ln(m_{3/2}))^q$ with $q >0$. This implies that the tilt in the distribution favouring lower values of $m_{3/2}$ 
also corresponds to the same for the scale of supersymmetry breaking in the visible sector. 

\item{\bf{Comparison with data:}}      Statistical distributions of vacua have been used to confront classes of string vacua with data (see e.g. \cite{Baer:2019xww, Baer:2020dri, Baer:2021vrk, Baer:2021tta, Baer:2021zbj, Baer:2022wxe, Baer:2022naw, Baer:2022qqr}) by examining the implications of the distributions of UV
parameters for low energy observables. These studies have focused on power-law and logarithmic distributions with a preference for high scale supersymmetry. It will be interesting to carry out similar analysis with our results given that the tilt in favour of low scale supersymmetry
is likely to help to make contact with observations or even to generate some tension with data. Of course, this line of study assumes that the distributions computed  by studying the distribution of vacua should be translated to  distributions for predictions for experiments\footnote{Emergence of selection principle(s) in the space of vacua (which can arise from early universe cosmology) can invalidate this assumption.}. 
  
\item {\bf{Multiple throats:}} The simplest generalisation of our setup is to consider multiple throats (this was studied in the context of supersymmetry breaking in \cite{Susskind:2004uv}). We examine it to understand the effect of multiple supersymmetry breaking sectors. 

Let us consider $n$ throats, each with a single anti-brane. Taking $n$ to be small,
so that the distribution in the throat sector factorises,\footnote{ Factorisation will
break down if there are many throats. At present, we do not have the tools to compute the distributions in such cases.} we have
\begin{equation}
\label{TwothD}
  \dd{\mathcal{N}_{\rm th}} \propto \prod_{i=1}^n {1 \over {y_i} (\ln y_i)^2 } \prod_{i=1}^n \dd{y_{i}}\,.
\end{equation}
The uplift term is given by
\be
 V_{\rm up} = {1 \over 2 s \tau^{2} } \sum_{i=1}^n y_i \, .
\ee
It is useful to make a series of variable changes: $y_i = w_i^{2}$, then to spherical coordinates in $w_i$ (with $r^{2} = \sum_{i=1}^n (w_i)^2$)
and angular variables ($\theta_j$, $j =1 ,...,(n-1)$)  and finally
$\tilde{r} =  r^{2}$. With this, the distribution \pref{TwothD} takes the form
\begin{equation}
\label{TwothD}
  \dd{\mathcal{N}_{\rm th}} \propto  {1 \over {2 \tilde{r} g(\theta_i)  \prod_{i=1}^n (\ln w_i)^2} }  \dd{\Omega_{n-1}} \dd{\tilde{r}},
\end{equation}
where $g(\theta_i)$ is a function of the angular variable. In these coordinates, the uplift term is given by
\be
 V_{\rm up} = {\tilde{r} \over 2 s \tau^{2} }\, ,
\ee
it is independent of the angular coordinates and has exactly the same form as the single variable case.
To obtain the joint distribution of the gravitino mass and the cosmological constant, consider the full distribution function
\be
  \dd{\mathcal{N}} =  { \eta'\,W_0 \over s^{2}} \dd{S} \dd{W_0}   \dd{\mathcal{N}_{\rm th}} \, ,
\ee
and make the variable change 
$ (W_0, \tilde{r}, s, C_0, \theta_i) \to (W_0, \Lambda, m_{3/2}, C_0, \theta_i)$ for LVS and $(W_0, \tilde{r}, S, \theta_i) \to (m_{3/2}, \Lambda, S, \theta_i)$ for KKLT. Since the uplift term has the same functional form as in the single variable case, also the functional form of the Jacobian is the same as in the single variable
case. As a result, the $\tilde{r}$ dependence of the density is  similar to the $y$ dependence in the single variable case. Thus in the $\Lambda \to 0$ limit, the functional dependence on $m_{3/2}$ is the same as in the single throat case (up to logarithmic terms). We conclude that the distribution function of throats is such
that adding multiple sectors (but small in number) preserves the tilt in the distribution function in favour of lower values of $m_{3/2}$.

\item{\bf{Sensitivity to uplift:}} The tilt in the distribution for $m_{3/2}$ towards lower values is tied to the distribution for throat hierarchies.
In particular, the $y^{-1}$ factor in the distribution for hierarchies plays a crucial role.   Thus our result should
not be taken as providing  the general picture for the distribution of $m_{3/2}$ and the cosmological constant.
In order to gain a general understanding, the distribution has to be studied for various uplift mechanisms (see e.g. \cite{Burgess:2003ic, Westphal:2006tn,  Cicoli:2012fh, Gallego:2017dvd, Antoniadis:2018hqy, Cicoli:2015ylx}). To gain a general understanding of the joint distribution of the supersymmetry breaking and the cosmological constant scales in the whole flux landscape, one should then be able to determine the relative abundance of vacua characterised by different uplifting mechanisms. Our work represents just the first step forward in this direction. We leave this important direction for future work.

\item{\bf{Quantum corrections:}} The distributions that we have derived make use of the Wilsonian effective field theory obtained at the high compactification scale after integrating out stringy and Kaluza-Klein states. Physical quantities will be however affected by low energy quantum loops. These corrections can be \emph{large} for the cosmological constant, i.e. ${\mathcal{O}}(m_{3/2}^{4})$, but the conditions used  to obtain the small $\Lambda$ limit
of the distributions, i.e. $\Lambda \ll m_{3/2}^2$ in KKLT and $\Lambda \ll m_{3/2}^{3}$ in LVS, are stable against such corrections. Hence the form of the distribution functions obtained, is expected
to be stable against the incorporation of quantum effects.
\end{itemize}

\section*{Acknowledgements}

AM is supported in part by the SERB, DST, Government of India by the grant MTR/2019/000267. The work of K. Sinha is supported in part by DOE Grant desc0009956.

\bibliographystyle{JHEP}
\bibliography{references}

\end{document}